\newcount\style \style=2
\ifodd\style
 \documentstyle[aas2pp4]{article}
  \else
   \documentstyle[12pt,aasms]{article}
\fi 

\begin{document}
\setlength{\baselineskip}{12pt}

\newcommand\bb[1] {   \mbox{\boldmath{$#1$}}  }

    \def\bv { {\bf v} }
    \def\vv { {\bf v} }
\newcommand\del{\bb{\nabla}}
\newcommand\bcdot{\bb{\cdot}}
\newcommand\btimes{\bb{\times}}
    \def\dd{\partial}
    \def\tilde{\widetilde}
    \def\etal{et al.}
    \def\eg{e.g. }
    \def\etc{{\it etc.}}
    \def\ie{i.e.}
    \def\beq{ \begin{equation} }
    \def\eeq{ \end{equation} }
    \def\spose#1{\hbox to 0pt{#1\hss}} 
    \def\ltsim{\mathrel{\spose{\lower.5ex\hbox{$\mathchar"218$}}
	 \raise.4ex\hbox{$\mathchar"13C$}}}

\def\tilde{\widetilde}

\long\def\Ignore#1{\relax}

\title{On the Dynamical Foundations of $\bb{\alpha}$ Disks.}

\author{Steven A.\ Balbus}
\affil{Virginia Institute of Theoretical Astronomy, Department of
Astronomy, University of Virginia, Charlottesville, VA 22903-0818}
\affil{sb@virginia.edu}

\smallskip
\author{John C.\ B.\ Papaloizou}
\affil{Astronomy Unit, School of Mathematical Sciences, Queen Mary and
Westfield College, University of London, Mile End Road, London E1 4NS, UK}
\affil{J.C.B.Papaloizou@qmw.ac.uk}

\vskip 2 truein
\centerline{Submitted to the {\sl Astrophysical Journal.}}
\vskip 2 truein

\begin{abstract}
The dynamical foundations of $\alpha$ disk models are described.  At
the heart of the viscous formalism of accretion disk models are
correlations in the fluctuating components of the disk velocity,
magnetic field, and gravitational potential.  We relate these
correlations to the large scale mean flow dynamics used in
phenomenological viscous disk models.  MHD turbulence readily lends
itself to the $\alpha$ formalsim, but transport by self-gravity does
not.  Nonlocal transport is an intrinsic property of turbulent
self-gravitating disks, which in general cannot be captured by an
$\alpha$ model.  Local energy dissipation and $\alpha$-like behavior
can be re-established if the pattern speeds associated with the
amplitudes of an azimuthal Fourier decomposition of the turbulence are
everywhere close to the local rotation frequency.  In this situation,
global wave transport must be absent.  Shearing box simulations, which
employ boundary conditions forcing  local behavior, are probably not an
adequate tool for modeling the behavior of self-gravitating disks.  As
a matter of principle, it is possible that disks which hover near the
edge of gravitational stability may behave in accord with a local
$\alpha$ model, but global simulations performed to date suggest
matters are not this simple.


\end{abstract}

\section{Introduction}

For many years, the principle uncertainty and greatest impediment for
the development of accretion disk theory was an understanding of the
origin of turbulent transport.  In their classic paper, Shakura \&
Sunyaev (1973) made the physically reasonable and enormously productive {\it
ansatz\/} that, whatever the underlying cause for its existence, the
turbulent stress tensor $T_{ij}$ scaled with the local gas pressure
$P$.  They denoted the constant of proportionality as $\alpha$, and the
``$\alpha$ disk'' moniker has since become synonymous with the standard
disk model.   Despite the ongoing development of
increasingly sophisticated large scale numerical models, $\alpha$ disk
modeling still remains the central link between theory and
observations, the cornerstone of accretion disk phenomenology.

In the last several years, a promising candidate has emerged as the
physical basis for $\alpha$ disk models.  This is the
magneto-rotational (``Balbus-Hawley'') MHD instability (Balbus \&
Hawley 1998, and references therein).  A large and ever-increasing body
of numerical simulations leaves little doubt that this instability
leads to the turbulent enhancement of angular momentum  transport
within an accretion disk.  What has been lacking, however, is a
systematic explanation of how turbulence may or may not lead to the
phenomenological $\alpha$ disk equations which have been in use for
many decades.   Since this approach has been (and continues to be) the
link between accretion disk theory and observations, a better
understanding of the $\alpha$ formalism is clearly desirable.  The
present paper seeks to fill this role.

The dynamical foundations of viscous disk theory have always been
somewhat fuzzy, and the benefits of sharpening our understanding are
numerous.  For example, it is desirable to clarify which results of
$\alpha$ disk phenomenology are truly fundamental, and which have more
limited domains of applicability.  The nature of time-dependent
turbulent transport could be more fully elucidated: in what sense is it
equivalent to a viscous stress?   Most interestingly, by introducing
the intermediate integral scales of turbulence into the investigation,
an enormously richer class of physical problem emerges.  Classical
$\alpha$ disk theory addresses mean flow (macroscopic) dynamics,
subsuming all integral scale structure into a viscous stress tensor.
In this approach, macroscopic disk structure is coupled directly to the
dissipative scales.  One cannot begin to answer such questions as
whether the turbulence is self-maintaining, or whether the transport is
local or global; everything is simply prescribed.  Finally, there are
important questions facing disk modelers in nonmagnetized disks, such
as protostellar disks on scales larger than a few AU.  Is it sensible
to model such regions with an $\alpha$ viscosity?  Are disks which evolve
under the influence of self-gravity $\alpha$ disks?  What distinguishes
turbulence well-modeled by $\alpha$ viscosity,  from turbulence which
is not?  These and related questions form the focus of this paper.

An overview of this paper is as follows.  In \S 2, we present a review
of classical viscous accretion theory.  Although this material is
well-known, we revisit it with a renewed attention upon how the viscous
stress appears in the angular momentum and energy fluxes.  This becomes
a benchmark for the turbulent theories discussed in \S\S\ 3 and 4.  In
\S 3\, we show that MHD turbulence acts very much along the lines
of classical viscous theory, in both steady-state as well as
evolutionary disk models.  In \S 4, it is shown the turbulent transport
arising from self-gravity is not, in general, compatible with a viscous
formalism.  We discuss the physical basis for this behavior, and show
that there are limiting cases of restrictive generality, which {\em
are\/} compatible with $\alpha$ disk theory.  Finally \S 5 summarizes
our findings.

\section {Preliminaries}
\subsection{Classical Viscous Disk Theory}

We begin with a brief review of classical viscous disk theory
(Lynden-Bell \& Pringle 1974, Pringle 1981).  In Cartesian
coordinates, with $x$, $y$, $z$ represented by dummy indices $i$, $j$,
$k$, the viscous stress tensor takes the form (Landau \& Lifschitz
1959):
\beq
\sigma_{ij} =  \eta(\dd_j v_i + \dd_i v_j
-{2\over3}\delta_{ij}\dd_k v_k).
\eeq
We use the standard notational convention of $\dd_i$ denoting the
partial derivative with respect to spatial coordinate $i$, and
summation over repeated indices is implied unless stated otherwise.
$v_i$ is the $i$th component of the velocity vector, and $\eta$ is the
dynamical viscosity.  To the extent that the fluid behaves
incompressibly, we may ignore the divergence term,  a standard and
generally well-justified procedure for the class of turbulence we wish
to consider here.  The idea of viscous disk theory is to regard the
effects of turbulence as greatly enhancing the magnitude of $\eta$
beyond its microscopic value, and doing nothing else.

The dynamical equations are mass conservation,
\beq\label{mass}
\dd_t \rho + \dd_j (\rho v_j) = 0,
\eeq
and the equation of motion
\beq\label{mom}
\rho (\dd_t v_i + v_j\dd_j v_i) = -\dd_j(P\delta_{ij} -\sigma_{ij})
-\rho\dd_i\Phi,
\eeq
($\delta_{ij}$ is the Kronecker delta function) which may also be written
\beq\label{momflx}
\dd_t(\rho v_i) + \dd_j(\rho v_i v_j +P\delta_{ij} -\sigma_{ij}) =
-\rho \dd_i \Phi,
\eeq
an explicit statement of momentum conservation. (Our notation is again
standard: $\rho$ is the mass density, $\Phi$ is the gravitational
potential, $P$ is the gas pressure.)  At this stage, we assume that
$\Phi$ is an imposed central disk potential; self-gravity is 
considered in \S 4.  In viscous disk theories, momentum transport ---
or more usefully, angular momentum transport---is the task of
$\sigma_{ij}$.  If the differential rotation rate decreases with
increasing radius,
viscosity transports angular momentum outward.

Multiplying eq.~[\ref{mom}] by $v_i$, integrating terms by parts,
using mass conservation,
and finally summing over $i$, leads to a mechanical energy equation
\beq\label{mech}
\dd_t (\rho v^2/2 + \rho \Phi) + \dd_j (\rho v^2 v_j/2+\rho \Phi +Pv_j
-v_i\sigma_{ij}) = P\dd_j v_j - (\dd_jv_i)\sigma_{ij},
\eeq
where
\beq
v^2 = v_iv_i
\eeq
The right hand side of eq.~[\ref{mech}] represents work done on the
fluid and heating of the fluid, respectively.  The presence of work and
heating terms
links disk mechanics with thermodynamics.   Here, we wish to highlight
the dual role of $\sigma_{ij}$: in eq.~[\ref{mom}],
it is a term in the transport of mechanical
energy flux; coupled in eq.~[\ref{mech}] to the strain $\dd_jv_i$, it is a
mechanical energy loss term.  In viscous disk models, the latter is,
of course, the origin of
accretion disk luminosity.  Since $\sigma_{ij}$ is a symmetric
tensor,
\beq\label{endis}
(\dd_jv_i)\sigma_{ij} = (1/2) (\dd_jv_i+\dd_iv_j)\sigma_{ij}=
(1/2) \eta^{-1} \sigma_{ij}\sigma_{ij} >0
\eeq
so the dissipated energy ultimately radiated is necessarily positive definite
for incompressible turbulence (Landau \& Lifschitz 1959).  

In cylindrical coordinates $(R, \phi, z)$ the azimuthal equation of motion
for an axisymmetric viscous disk expresses angular momentum
conservation, and takes the explicit form
\beq\label{angmom}
\dd_t(\rho R v_\phi) + \mbox{\boldmath {$\nabla$}}\cdot
(\rho R v_\phi {\bf v} 
-\eta R^2 \bb{\nabla}\Omega) =0.
\eeq
The rotational velocity $v_\phi$ is Keplerian,
\beq
v_\phi^2 = {GM\over R}
\eeq
where $M$ is the central mass, and 
\beq
R \Omega = v_\phi.
\eeq
In the simplest form of viscous disk theory, we ignore the vertical
structure, treating the disk as flat, set $v_\phi = R\Omega$,
and assume axisymmetry.
That is, we use the height integrated
form of the mass, angular momentum, and energy equations.
With $\Sigma$ denoting
the disk column density, mass conservation becomes
\beq\label{pringmass}
{\dd \Sigma\over \dd t} + {1\over R}{\dd R\Sigma {v_R}\over \dd R}
=0, 
\eeq
while angular momentum conservation follows immediately
from eq.~[\ref{angmom}]:
\beq\label{pringmom}
{\dd\over \dd t}(\Sigma R^2\Omega ) + {1\over R} {\dd \over\dd R}
(\Sigma R^3 \Omega v_R -\nu \Sigma R^3 {d\Omega\over dR}) =0,
\eeq
where $\nu$ is the kinematic viscosity, $\eta\equiv\rho\nu$.
The energy dissipated per unit
area of the disk $Q_e$ (``emissivity''),
is found from eq.~[\ref{endis}] to be
\beq\label{thirteen} Q_e = (9/8)\nu\Sigma \Omega^2.\eeq

Under steady state conditions, the mass flux ${\dot M} \equiv - 2\pi R
\Sigma v_R $ and the angular momentum flux must both be constant.
Assuming that the viscous stress vanishes at the inner edge of the disk
$R_0$ leads to the relation
\beq
{\dot M}  \left( 1 - \left(R_0\over R\right)^{1/2} \right)=3\pi \nu \Sigma.
\eeq
The turbulent paramater $\nu\Sigma$ is severely restricted by this relation,
and can be eliminated in favor of ${\dot M}$ in the expression for the 
emissivity, leading to (Pringle 1981):
\beq\label{qmdot}
Q_e = {3GM{\dot M}\over 8\pi R^3}\left(  1 - \left(R_0\over R\right)^{1/2}
\right).
\eeq This is the classical $(Q_e , {\dot M})$ relationship, which leads to a
surface emission temperature profile $T_{eff}(R) \propto R^{-3/4}$.
Its utility lies in the absence of an explicit viscosity term, which has
been eliminated by the requirements of constant angular momentum and
mass flux.

We may drop the assumption of time steady conditions, forgoing a functional
restriction for $\nu \Sigma$ in the process.  Using eq.~[\ref{pringmass}]
in eq.~[\ref{pringmom}], leads to a generalized mass accretion  
formula:
\beq\label{gmaf}
\Sigma v_R R = - 3 R^{1/2} {\dd\ \over\dd R} (\nu \Sigma R^2 \Omega).
\eeq
This in turn may be used back in
eq.~[\ref{pringmass}], yielding an equation for $\Sigma$ in
terms of $\nu$ (Lynden-Bell \& Pringle 1974):
\beq\label{evol}
{\dd\Sigma\over \dd t} = {3\over R} {\dd\ \over\dd R}\left[  R^{1/2} {\dd\ \over
\dd R}\left(\nu R^{1/2}\Sigma\right)\right]
\eeq
This is the classical evolutionary equation commonly used in accretion
disk modeling.  It requires an {\it a priori} specification of
the functional dependence of $\nu$ to be useful, and leads to
diffusive behavior (disk spreading) in eruptive systems.
Ultimately mass is transported inwards and angular momentum
outwards:  all of the latter is in a vanishingly small component of
the former.

\section {Magnetohydrodynamical Turbulence.}

The fundamental assumption underlying essentially all phenomenological
modeling of turbulent disks is the following: it makes sense to use a
two scale approach to mathematically represent disk attributes of
astrophysical interest.  With the possible exception of ``flickering''
in CV systems (e.g., Welsh, Wood, \& Horne 1996), observational data is
assumed to involve length and time scales which are larger than the
characteristic ``eddy turnover'' scales of the turbulence.  One works
with averages that are assumed to be well-defined, and to represent the
large-scale properties of the disk, much as classical dynamo theory
(Krause \& R\"adler 1980) represents mean fields and mean helicity.
While this statement probably does not strike the reader as startling
or controversial, it masks subtleties.  The assumuption has remarkably
restrictive consequences.

The problem is that power spectrum of almost all nondissipative
quantities (including the stress tensor itself) is dominated by the
largest scales of the turbulence, in this case, the disk scale height
and rotation period.  The implicit averaging must be on scales large
compared to the scale height but small compared with the radius, and
large compared with the orbital period but small compared with viscous
and thermal time scales.  There must be an asymptotic domain where these
scales are cleanly separated, so that a computed radial disk profile is
insensitive to the averaging procedure.  Let us assume this is the case
and pursue its consequences.

\def\B {{\bf B}}
The dynamical equation of motion in the presence of a magnetic field
$\B$ is 
\beq\label{eqomo}
\rho {\dd\bb{v} \over \dd t} + (\rho \bb{v}\bcdot\del)\bb{v} = -\del\left(
P + {B^2\over 8 \pi} \right)-\rho \del \Phi +
\left( {\B\over
4\pi}\bcdot \del\right)\B +\eta_V \nabla^2\bb{v}.
\eeq
We have denoted the viscosity as $\eta_V$ to distinguish it from the
resistivity associated with the magnetic field, which we will denote as
$\eta_B$.   We have dropped terms proportional to $\del\bcdot\bb{v}$ in
the viscous term.  The azimuthal component of this equation can be
written in a form which expresses angular momentum conservation:
\beq\label{angmomB}
{\dd\ \over \dd t}(\rho R v_\phi) + \del\bcdot R \left[ \rho v_\phi\bb{v}
- {B_\phi \over4\pi} \bb{B_p} +\left(P+{B_p^2\over8\pi}\right)
\bb{{\hat e}_\phi} - \eta_V R^2\del(v_\phi/R)\right]
=0,
\eeq
where the subscript $\bb{p}$ denotes a poloidal vector component.  

We now separate the circular motion $R\Omega$ from the noncircular
motion ${\bf u}$, treating the latter as a fluctuating quantity,
though not necessarily with vanishing mean.  We have
\beq
\bb{v} = R\Omega \bb{{\hat e}_\phi} +\bb{ u}.
\eeq
Although mean drift velocities may be present (the disk must accrete),
we assume that such motions are small compared with fluctuations
amplitudes,
\beq
|\langle \bb {u} \rangle |^2 \ll \langle u^2 \rangle,
\eeq
where the angle brackets denote a suitable average, discussed below.
Furthermore, the direct contribution to the angular momentum flux from 
the microscopic viscosity $\eta_V$ is generally negligible.  (This is why
turbulence is necessary!)  We may drop this term.

Substituting for $\bb{v}$ in eq.~(\ref{angmomB}), and averaging over
azimuth, denoting such means as $\langle\rangle_\phi$.  This gives
\beq\label{aa}
{\dd\ \over \dd t}\langle \rho R^2 \Omega\rangle_\phi +
\del\bcdot R \left[ \langle \rho R\Omega\bb{u_p}\rangle_\phi +
\bb{T}\right] =0,
\eeq
where $\bb{T}$ is the poloidal stress tensor
\beq
\bb{T} = \langle \rho u_\phi \bb{u_p} - B_\phi \bb{B_p}/4\pi\rangle_\phi
\eeq
We have dropped the $\rho u_\phi$ term in comparison with $\rho R \Omega$ in
the leading time derivative.   Henceforth, we will use the Alfven
velocity
\beq
\bb{u_A} \equiv { \bb{B}\over \sqrt{4 \pi\rho}}
\eeq
in favor of the magnetic field vector $\bb{B}$.

\subsection {MHD Turbulence and Viscous Disk Theory}

To make contact with the classical viscous disk
theory of the previous section we need to integrate eq.~(\ref{aa}) over
$z$, and to assume that surface terms can be dropped.  Furthermore, we wish
to regard height-integrated, azimuthal averaged flow quantities as smooth
functions of $R$.  As indicated above, this
also implies some sort of radial smoothing---an average over a volume
small compared with $R$, but larger than a disk scale height.  We follow
the notation of Balbus \& Hawley (1998), and define the density weighted mean of
flow attribute $X$ to be
\beq
\langle X\rangle_\rho \equiv {1\over 2\pi\Sigma \Delta R}\int^{\infty}_{-\infty}\int^{R+\Delta R/2}
_{R-\Delta R/2} \int^{2\pi}_0 \rho X
\> d\phi\, dR\,  dz \label{average}
\eeq
where $\Sigma$ is the integrated  and similarly radially averaged disk column density.  The radial
component of $\bb{T}/\rho $ resulting from this operation will be denoted simply as
$W_{R\phi}$.  Angular momentum conservation becomes
\beq\label{ab}
{\dd\ \over \dd t}\langle \Sigma R^2 \Omega\rangle_\rho +
{1\over R}{\dd \over\dd R}\left( R^3\Omega \Sigma\langle u_R\rangle_\rho
+R^2\Sigma W_{R\phi}\right) = 0
\eeq
Mass conservation follows straightforwardly from integrating the
fundamental equation, and leads to a form essentially identical to
eq.~[\ref{pringmass}]:
\beq\label{avgmass}
{\dd\Sigma \over \dd t} + {1\over R} {\dd (R \Sigma \langle
u_R\rangle_\rho) \over \dd R} = 0 
\eeq
Using this in eq.~(\ref{ab}) gives a general formula for the mass
accretion rate
\beq\label{ac}
\Sigma \langle u_R \rangle_\rho = - {1\over R (R^2 \Omega)'} {\dd\
\over\dd R} (\Sigma R^2 W_{R\phi}),
\eeq
where the prime $'$ denotes differentiation with respect to $R$.
Combining eqs.~(\ref{ac}) and (\ref{avgmass}) gives us the analogue to
eq.~(\ref{evol}) in a turbulent disk, for any angular momentum profile:
\beq\label{turbevol}
{\dd\Sigma\over \dd t} = {1\over R} {\dd\ \over \dd R} {1\over 
(R^2\Omega)'} {\dd\ \over \dd R} (\Sigma R^2 W_{R\phi}).
\eeq

Since $W_{R\phi}$ of not known {\em a priori,} in practical terms,
eq.~(\ref{turbevol}) represents only a marginal improvement on the
phenomenological equation (\ref{evol}).  But we may see that
``viscous'' evolution does not require the explicit adoption of a
viscous stress tensor.  Any disk in which $u_R$ and $u_\phi$ (and
$u_{AR}$ and $u_{A\phi}$) are positively correlated must behave
similarly, with the caveat that the correlation tensor must be a
locally defined quantity.

We have discussed thus far only the dynamics of the turbulence.  Once
$W_{R\phi}$ is known, and it depends primarily on correlations on the
largest turbulent scales, the disk evolution may be directly calculated
by equation (\ref{turbevol}).  Classical viscous disk theory also
addresses the energetics.  Since viscosity is the agent of transport,
there must be dissipation as well.  The energy is directly thermalized
from its free source in the differential rotation, down to thermal
scales.

In a turbulent disk, matters are more complex.  Energy cascades from
the differential rotation to the scales of the largest fluctuations,
thence to the integral self-similar scales, and finally to the
dissipative Kolmogorov scale (which may be set by resistivity rather
than viscosity).  In a steady state disk, we expect that the rate at
which energy is extracted from the differential rotation, which may be
easily calculated in terms of the stress tensor, to be equal to the
rate at which it is thermalized, which would otherwise not be directly
calculable.  The upshot of this is that the steady state turbulent
disks behaves viscously in their energetics as well in their dynamics.
But classical viscous theory makes a stronger assumption by its very
nature: the rate of thermalization of the free energy of differential
rotation is the same in both steady and evolutionary models.  This {\it
may} be true in an evolving turbulent disk, but it is not obviously
true.  It depends upon whether the cascade is efficient.  Fortunately,
we shall see that this question may be directly and quantitatively
answered within the stress tensor formalism we are using.

The evolution of the magnetic field in a plasma with resistivity
$\eta_B$ is given by
\beq \label{ind}
{\dd\bb{B}\over \dd t} = \del\btimes\left( \bb{v} \btimes \bb{B} -
\eta_B \del\btimes\bb{B}\right).
\eeq
The energy of mechanical energy equation is obtained by dotting
eq.~(\ref{eqomo}) with $\bb{v}$, dotting eq.~(\ref{ind}) with 
$\bb{B}$, and combining the two.  After some simplification
(e.g.,  Balbus \& Hawley 1998), we arrive at
\beq \label{mechen}
{\dd\ \over\dd t} \left( {1\over2} \rho v^2 + \rho\Phi+ {B^2\over8\pi}\right)
+ \del\bcdot[\ ] = 
P \del\cdot\bb{v} - \eta_V(\dd_iv_j)(\dd_iv_j)
- {\eta_B\over4\pi} |\del\btimes\bb{B}|^2
\eeq
We have dropped the term $-\eta_V|\del\bcdot\bb{v}|^2$ on the right
hand side since in a turbulent disk it is generally small compared with
the other viscous term.  The right hand pressure term, though
also proportional to $\del\bcdot\bb{v}$, cannot be dropped; 
as noted in \S 2, it
represents a link to the internal energy of the disk via the first law
of thermodynamics.  The unwritten energy flux in square brackets is 
\beq
\bb{v}\left( {1\over2}\rho v^2 +\rho\Phi+P\right) + {\bb{B}\over4\pi}\btimes
(\bb{v}\btimes\bb{B})
\eeq
where, as before, we have not included the transport due to explicit
viscosity or resistivity.

The energy flux is more complex than its angular momentum counterpart,
but can be greatly simplified by retaining only leading terms in a $u\ll R\Omega$
expansion.  When averaged as before, the radial energy flux is
\beq\label{eflux}
\Sigma ( {1\over2} R^2 \Omega^2 +\Phi)
\langle u_R\rangle_\rho  + \Sigma R\Omega W_{R\phi},
\eeq
which may be compared with the radial angular momentum flux of eq.~(\ref{ab}),
\beq\label{momflux}
\Sigma R^2 \Omega \langle u_R\rangle_\rho + \Sigma R W_{R\phi}.
\eeq
(Note that to effect the height integration, we have assumed that the
magnetic field is force-free above the disk.  This is a physically
reasonable assumption, but one that is less than general.
The energy dissipation rate is, however, indifferent to
the presence or absence of surface terms resulting from vertical integrations,
since the ignored vertical
fluxes would contribute nothing to the local mechanical energy losses
in the disk.)
The key point is to observe that the only turbulence parameters entering
into either the angular momentum or energy radial fluxes are $\langle
u_R\rangle_\rho$ and $W_{R\phi}$.  {\em This is the essence of an
$\alpha$ disk.}

The issue is most clear for a steady model (Balbus \& Hawley 1998).
In this case the accretion rate 
\beq
{\dot M} \equiv - 2 \pi  \Sigma R \langle u_R\rangle_\rho
\eeq
is constant, and if the angular momentum flux at the inner edge of the disk
$R_0$ is vanishingly small, then 
\beq
W_{R\phi} = { {\dot M} \Omega \over 2\pi \Sigma}\left[ 1 - \left(R_0\over
R\right)^{1/2} \right],
\eeq
Now, one cannot calculate directly the thermalization losses, since
the small scale gradients are not known.  But one can calculate
the divergence of the large scale flux, since the spatial
dependence of $W_{R\phi}$ is determined by angular momentum
conservation.  This must be the small scale dissipation rate.
We find
\beq\label{abc}
Q_e = -\Sigma W_{R\phi} {d\Omega\over d\ln R},
\eeq
which is precisely the analogue of eq.~(\ref{thirteen}) if $W_{R\phi}$
is replaced by a large scale viscous stress.  (Note that $Q_e > 0$.)
This result can also be obtained directly from the energy equation for
the $u$ fluctuations themselves, by demanding that sources and sinks
balance in steady state (Balbus \& Hawley 1998).

More is required, however.  In viscous models, the thermalization
rate is given by eq.~(\ref{thirteen}) whether steady conditions prevail or not.
The question before us is whether the thermalization rate (\ref{abc})
is just as general: does it hold when ${\dot M}$ is not constant and
when the energy density of the disk changes with time?  We now show
that it does.

First, let us recall the fundamental relations for the angular and
epicyclic frequencies,
\beq\label{fund}
\Omega^2 = {1\over R} {\dd\Phi\over \dd R}, \quad \kappa^2 = {1\over
R^3} {\dd\ \over\dd R} (R^4\Omega^2) = {1\over R^3}{\dd\ \over \dd R}
(R^3\Phi'),
\eeq
as well as the specific energy,
\beq
{1\over 2} R^2\Omega^2 +\Phi = {1\over 2R} {\dd\ \over\dd R}(R^2\Phi).
\eeq
Thus, the energy flux of eq.~(\ref{eflux}) becomes
\beq
{1\over 2R} \Sigma \langle u_R\rangle_\rho {\dd\ \over\dd R}(R^2\Phi)
+ \Sigma R\Omega W_{R\phi}.
\eeq
We may substitute for $\langle u_R\rangle_\rho$ using equation
 (\ref{ac}).  This gives an energy flux of
\beq
-\Omega {(R^2 \Phi)'\over (R^3 \Phi')'} {\dd\ \over \dd R} (\Sigma R^2
W_{R\phi}) + \Sigma R \Omega W_{R\phi}.
\eeq

The quantity of interest is the heating rate $Q_e$ due to turbulent
dissipation per unit area, and it is given by vertically integrating  
the left-hand side of eq.~(\ref{mechen}).
Making use of the above it may be written in the form
\beq\label{qwe} -Q_e=
{1\over 2R}  {\dd R^2\Phi_s\over\dd R} {\dd\Sigma \over \dd t} 
+ {1\over R} {\dd\ \over \dd R} R
\left[- \Omega{(R^2 \Phi)'\over (R^3 \Phi')'}{\dd\ \over \dd R}
(\Sigma R^2 W_{R\phi}) +
\Sigma R\Omega W_{R\phi}\right].
\eeq
If we now use eq.~(\ref{turbevol}) for $\dd\Sigma/\dd t$ in eq.~(\ref{qwe}),
we obtain
\begin{eqnarray}
-Q_e  &  =  &
{1\over R^2}
{\dd(R^2\Phi) \over\dd R} \cdot
{\dd\ \over \dd R}\left[ {R^2\Omega\over(R^3\Phi')'}
{\dd\ \over \dd R}(\Sigma R^2 W_{R\phi})\right]\nonumber \\
 &  & \quad - {1\over R}{\dd\ \over \dd R}  \left[R\Omega {(R^2 \Phi)'\over (R^3
\Phi')'} {\dd\ \over \dd R}(\Sigma R^2 W_{R\phi}) - \Sigma R^2\Omega
W_{R\phi}\right].
\end{eqnarray}
This unwieldy formula immediately simplifies to 
\beq -Q_e =
 - {R\Omega\over (R^3\Phi')'} \cdot {\dd\ \over \dd R}(\Sigma R^2
 W_{R\phi}) \cdot {\dd\ \over \dd R}{1\over R} {\dd\ \over \dd R}(R^2\Phi)
 + {1\over R} {\dd\ \over \dd R} (\Sigma R^2 \Omega W_{R\phi}).
 \eeq
Furthermore, for any function $\Phi$, the following identity is easily
verified,
\beq
( R^{-1}( R^2\Phi)')' = R^{-2} (R^3\Phi')',
\eeq
leading to a complete collapse of our expression down to the
single term
\beq -Q_e=
\Sigma W_{R\phi} {d\Omega\over d\ln R},
\eeq
which is the desired result.  To leading order in the turbulent
fluctuation amplitudes, the  thermalization rate per unit area  of a magnetized
disk is given by the above, whether the disk is evolving or in a steady
state.   This result is nicely compatible with classical viscous thin disk theory.
That not all disk turbulence is so easily subsumed 
will be seen in the next section.

\section {Self-Gravity}

Self gravitational forces can be important for galactic and
protostellar disks.  In its most extreme manifestation, self-gravity
can hold the disk together and cause substantial deviations from a Keplerian
rotation law.  But this requires a disk mass comparable to or in
excess of the central compact mass, and we will not consider this
limit.  Instead we focus on a more common situation in which the
local self-gravitating free fall time $\sim (G\rho)^{-1/2}$ is
comparable to or smaller than the $1/\Omega$.  There are  several
equivalent ways of expressing this condition, the classical Toomre
(1964)
Q criterion being the best known (Binney \& Tremaine 1987).
With $c_S$ denoting the sound speed, if
\beq
Q \equiv {\kappa c_S\over \pi G \Sigma} < 1,
\eeq
then local density perturbations are unstable to  gravitational
collapse in a thin disk. 
If we define the vertical scale height  $H$ by 
$cH=\Omega$ and the disk density by $\rho H = \Sigma$,
then for a Keplerian disk the $Q$ criterion becomes
\beq\label{q_ref}
{\Omega^2\over \pi G\rho} < 1,
\eeq
in rough agreement with our initial estimate, and we may avoid an
explicit reference to the disk temperature.

Self-gravity is obviously important in the formation stages of a galaxy
or a star, but it is also likely to be a key component in later
evolutionary stages, especially in the outer regions of the disk where
$\Omega^2/\rho$ is likely to be small.  We shall concentrate here on
the latter case, assuming a well-defined Keplerian disk is present,
with self-gravity causing small but critical departures from circular
flow.  The ratio of disk mass $M_d$ to central mass $M$ is found from
eq.~(\ref{q_ref}) to be of order $(H/QR)$,
so our appoximation is justified for thin disks.

Progress in the numerical modeling of disk systems has been impressive,
and sophisticated simulations are now possible, although investigators
understandably tend to want to explore the more dramatic behavior of
very massive disks.  One of the interesting question these modelers are
addressing is whether turbulence wrought by self-gravity is amenable to
a viscous diffusion treatment (e.g., Laughlin \& Royczyska 1996).  We
now examine this point.

\subsection{Dynamical and Energy Fluxes}

The self-gravity potential $\Phi_S$ satisfies the Poisson equation, most
conveniently written in the form
\beq
\rho = {1\over4\pi G} \dd_i\dd_i \Phi_S, 
\eeq
where the subscript $i$ (or $j,k$ below) denotes a Cartesian coordinate,
and the summation convention on repeated subscripts is used unless otherwise stated.  
The connection between the gravitational force and its associated stress
tensor was first made by Lynden-Bell \& Kalnajs (1972):
\beq
- \rho \dd_i\Phi_S = -{\dd_i \Phi_S \over4\pi G} \dd_j\dd_j \Phi_S  =
{1\over 4\pi G} \dd_j\left[ -(\dd_j\Phi_S)(\dd_i\Phi_S) + {\delta_{ij}\over
2} (\dd_k\Phi_S)(\dd_k\Phi_S) \right] \eeq
To keep both the gravitational and nongravitational components of the
stress tensor on an equal footing, define the velocities
\beq
\bb{u_G} \equiv {\bb{\nabla}\Phi_S\over \sqrt{4\pi G\rho}}.
\eeq
Then, in the presence of self-gravity and magnetic fields,
the $R\phi$ component of the stress tensor becomes
\beq\label{Wgrav}
W_{R\phi} = \langle u_R u_\phi +u_{GR}u_{G\phi} - u_{AR} u_{A\phi}
\rangle_\rho
\eeq
Equations (26) and (29), angular momentum conservation
and the disk evolution equation, continue to hold in precisely the
same form when self-gravity is present, if the stress tensor
$W_{R\phi}$ is amended simply as above.  Gravitational torques are
calculated formally in exactly the same way as turbulent and
magnetic torques.  When $Q$ is of order unity, the kinetic $u$ terms
and gravitational $u_G$ terms of $W_{R\phi}$ are comparable if
$u\sim | \bb{\nabla} \Phi_S|/2\Omega$---i.e., if the fluctuation velocities
are due to self-gravity impulses on a rotation time scale.

Consider next the energetics of self-gravity.  Is the volumetric
dissipation rate still given by equation (46), with the gravitationally
amended version of $W_{R\phi}$?  The answer is not in general, but
under some interesting conditions it is.  Let us see how this emerges.

We seek to write the expression ${\rho v_i\dd_i\Phi_S}$ in
conservation form: the time derivative of an energy density plus
the divergence of a flux.  We have,
\beq
\rho v_i \dd_i \Phi_S = \dd_i(\rho v_i \Phi_S) - \Phi_S \dd_i (\rho
v_i) = \dd_i(\rho v_i \Phi_S) + \dd_t(\rho\Phi_S) - \rho\dd_t\Phi_S
\eeq
where the last equality follows from mass conservation plus an
integration by parts.  There is no sign yet of the gravitational
stress tensor putting in an appearance, but the final term remains
dangling for the moment.  This may be written
\beq
\rho {\dd_t \Phi_S} = {1\over 4\pi G} (\dd_i\dd_i\Phi_S)(\dd_t
\Phi_S) = {1\over 4\pi G}\dd_i[(\dd_i\Phi_S)(\dd_t\Phi_S)]
-{1\over 8\pi G}\dd_t [(\dd_j\Phi_S)(\dd_j\Phi_S)].
\eeq
Returning to vector invariant notation,
\beq\label{tgb}
\rho\bb{v}\bcdot \bb{\nabla}\Phi_S = {\dd\ \over \dd t} (\rho \Phi_S
+{1\over 8\pi G} |\del\Phi_S|^2 ) + \del\bcdot \left( \rho\bb{v}\Phi_S -
{\del\Phi_S \over 4\pi G} {\dd\Phi_S\over \dd t}\right)
\eeq
There is some ambiguity as to whether one assigns terms to the
energy density or the flux.  For example, an equivalent formulation of
eq.~(\ref{tgb}) is
\beq
\rho\bb{v}\bcdot \bb{\nabla}\Phi_S = - {1\over8\pi G} {\dd\ \over \dd t}
|\del\Phi_S|^2 + \del\bcdot \left[ \rho\bb{v}\Phi_S -
{\del\Phi_S\over 4\pi G} {\dd\Phi_S\over\dd t}
+{1\over8\pi G}{\dd\ \over\dd t} \del\Phi_S^2\right].
\eeq
But there is no apportionment that of itself produces
an $R\phi$ component of the gravitational stress tensor.  
Since the energy flux is most readily interpretable in
eq.~(\ref{tgb}), we shall use this form of energy conservation in our discussion below.

The combination $-\del\Phi_S\cdot\dd\Phi_S/\dd t$ will be familiar to
students of acoustical theory (e.g., Lighthill 1978) where precisely
this form of energy flux is associated not with a gravitational
potential, but with the velocity potential of irrotational sound
waves.  In the acoustic case, this emerges from the ``$Pv$'' term in
the energy flux, a term which is third order in the fluctuation
amplitudes for incompressible turbulence, and therefore negligible.
Indeed, an important physical distinction between a disk  in which
there is a superposition of waves and a disk which is truly turbulent
is the dominance of the $W_{R\phi}$ term over the $Pv$ in the latter's
energy flux.

If waves were present in a turbulent disk, would this change the
relative dominance?  Not in a thin  non-self-gravitating
Keplerian disk with good $R\phi$
correlations in the stress tensor.  In a density wave, the pressure
contribution to the energy flux will be of order $u^2 c_S$ in the
velocities, whereas the stress tensor term is of order $\alpha R\Omega
c_S^2$.  Since $u^2\sim \alpha c_S^2$, the stress tensor contribution
will always be dominant (by a factor of $R/H$) in a thin disk.

The appearance of a second order contribution of the potential in the
energy flux suggests qualitatively new transport features in
self-gravitating disks.  In retrospect, the breakdown of the $\alpha$
formalism is perhaps not surprising.  Turbulence in hydrodynamical
shear flows or MHD disks arises because vorticity fields and magnetic
fields are ``ensnared'' by shear, and funnel this free energy into
fluctuations.  These fields may become ensnared because both are frozen
into their respective fluids.  Their evolution is entirely local, and
the vorticity and magnetic fields are governed by essentially identical
equations.  Gravitational fields are not frozen into the fluid, and we
should not expect local dissipation of its associated turbulence, which
is the inevitable consequence of an energy flux depending upon the
stress tensor and drift velocity, as may be seen in eq.~(33).
Self-gravity is generally a global phenomenon (its field equation is
elliptic), and one has no cause to expect a repetition of our earlier
magnetic success with local theory.

\subsection{The Local Limit}

If instead of the combination $\rho\bb{v}\Phi_S-
\del\Phi_S\,\dd_t\Phi_S$ appearing in the energy flux, the combination
$\Omega \del\Phi_S\cdot\dd_\phi \Phi_S$ emerged, we would be able to
construct a local model of the dissipation.  In this case, the
gravitational component of the stress tensor would couple energetically
precisely as magnetic and Reynolds stresses couple.  It will turn out that
the vanishing of {\em both\/} these terms corresponds to
\beq\label{co}  \left(
{\dd\ \over \dd t} + \Omega {\dd\ \over \dd\phi} \right)\Phi_S = 0,
\eeq
when the disturbances are analyzed in terms of WKB waves.  This is a
very revealing requirement, for it is just this condition that defines
the corotation resonance in linear density wave theory, and it is only
at this location that waves couple directly the disk (e.g., Goldreich
\& Tremaine 1979, hereafter GT).  It is quite natural, therefore, that
this condition reemerges as the requirement for gravitationally driven
energy stresses to be thermalized.

Let us examine the structure of the energy conservation equation
further.  We focus for simplicity upon an unmagnetized discs, and
assume that
$R\Omega \gg u_G$.  Denoting the volumetric mechanical energy losses
as $-\varepsilon$, the self-gravitational analogue to
equation (\ref{mechen}) becomes
\beq \label{mechengrav}
{\dd\ \over\dd t} \left[ \rho({v^2\over2} +\Phi+ \Phi_S) 
+{1\over 8\pi G} |\del\Phi_S|^2\right]
+ \del\bcdot \left[ \rho\bb{v}( {v^2\over2}+\Phi+\Phi_S ) - 
{\del\Phi_S\over 4\pi G} \left({\dd\Phi_S\over\dd t}\right) 
\right] = -\varepsilon.
\eeq
We have neglected
the pressure contribution to the energy flux (but see below).

We rewrite the rate of production of mechanical work as given by the left
hand side of (\ref{mechengrav}), separating the terms in a suggestive manner:
\beq \label{mechengrav1}
{\dd\ \over\dd t} {\cal E} 
+ \del\bcdot \bb{ {\cal F}} 
+\del\bcdot\left(\rho \Phi_S \bb{v}
-{\del\Phi_S\over 4\pi G} \> {D\Phi_S\over Dt} \right)
=
-\varepsilon,
\eeq
where the energy density ${\cal E}$ is
\beq
{\cal E} =  \rho({v^2\over2} +\Phi+\Phi_S) +{1\over 8\pi G}
|\del\Phi_S|^2,
\eeq
the local flux $\bb{{\cal F}}$ is
\beq
 \bb{{\cal F}}= \rho\bb{v}(\Phi +{v^2\over2})
+\Omega {\del\Phi_S\over 4\pi G} {\dd\Phi_S\over\dd \phi},
\eeq

and
\beq
{D\ \over Dt} \equiv {\dd\ \over \dd t} + \Omega {\dd \over \dd\phi}.
\eeq
The final terms on the left
side of the eq. (\ref{mechengrav1}) are ``anomalous'' from the point-of-view of
$\alpha$ disk theory, and the flux will henceforth be denoted as 
$\bb{F^*_E}$.
The radial component of $\bb{{\cal F}}$ is just given by eq.~(\ref{eflux}),
with $W_{R\phi}$ amended as in eq.~(\ref{Wgrav}).
Were $\bb{F^*_E}$ negligible, we would be lead directly to
an $\alpha$ disk model via the route we followed in \S\S\ 2 and 3.
However, since these terms may be of order $R\Omega u_G^2$ in the
velocities, they cannot be neglected.  Their physical
interpretation is discussed in the next section.

\subsection{Wave Fluxes in Self-Gravitating Disks}

To understand the role of the anomalous flux, it is helpful to study it
in the context of the simplest possible fluctuating
self-gravitating
disk model: WKB waves in a thin, pressureless disk.  (The inclusion
of pressure terms leads to a more complicated calculation, but with
precisely the same final conclusion.)
The waves have the canonical form 
$\exp(i\int^R kdx +im\phi - i\omega t)$, where
$k$ is the local radial wavenumber $m$ the azimuthal
wavenumber variable, and $\omega$ the fixed wave frequency,
and satisfy the dispersion relation 
\beq
(\omega- m\Omega)^2 = \kappa^2 - 2 \pi G \Sigma |k|.
\eeq
The potential $\Phi_S$ has the vertical spatial dependence $e^{-|kz|}$
out of the disk midplane (Lin \& Shu 1966, Binney \& Tremaine 1987).

The radial
anomalous energy flux, averaged over azimuth and integrated over height is
\begin{eqnarray}
 \left\langle \bb{F^*_E\cdot \hat{e}_R}\right\rangle & 
= &  \int^{\infty}_{- \infty} \left\langle
\rho  u_R \Phi_S -{1\over4\pi G}
{\dd\Phi_S\over \dd R}\left( {\dd\Phi_S\over \dd t} + \Omega {\dd\Phi_S\over \dd
\phi} \right) \right\rangle_\phi\, dz \nonumber\\
    & = &\int^{\infty}_{- \infty} \left\langle
    \rho  u_R \Phi_S \right\rangle_\phi\, dz
    - {1\over 4 \pi G}
    \left(\Omega - {\omega\over m} \right) \int^{\infty}_{- \infty} \left\langle
    {\dd\Phi_S\over \dd R} {\dd\Phi_S\over \dd \phi} \right\rangle_\phi\, dz
\end{eqnarray}
To do the first integral, we need to be able to express $u_R$ in terms of $\Phi_S$.
This relation may be read off directly
from eq.~(11) of GT:
\beq
u_R = {m\Omega-\omega\over 2\pi G \Sigma} \Phi_S(0)\ {\rm sgn}(k).
\eeq
where $\Phi_S(0)$ is the midplane ($z=0$) value of the potential.
Denoting the potential amplitude by
$\tilde{\Phi}_S$, and assuming $\rho(z) = \Sigma \delta(z)$,
the first integral is then
\beq
 \int^{\infty}_{- \infty} \left\langle \rho  u_R \Phi_S \right\rangle_\phi\, dz
=  {1\over 4 \pi G} (m\Omega - \omega ) {\tilde{\Phi}_S}^2 \ {\rm sgn}(k)
\eeq
The second integral may be evaluated by noting that the integrand depends
on $z$ as $\exp(-2|kz|)$.  This gives
\beq
 - {1\over 4 \pi G}
 \left(\Omega - {\omega\over m} \right) \int^{\infty}_{- \infty}
 \left\langle {\dd\Phi_S\over \dd R} {\dd\Phi_S\over \dd \phi} \right\rangle_\phi\,
	 dz
=
-{1\over 8 \pi G} (m\Omega - \omega) { \tilde{\Phi}_S}^2 \ {\rm sgn}(k).
\eeq
Thus,
\beq
\bb{F^*_E\cdot \hat{e}_R} =  - {1\over 8 \pi G} ( \omega - m\Omega )
{\tilde{\Phi}_S}^2 \ {\rm sgn}(k)
\eeq

The angular momentum flux is also to be found in GT (eq.~(30); note their
definition differs by a factor of $2\pi R$ from ours).  It is simply
\beq
F^*_J = - {m\over 8 \pi G} {\tilde{\Phi}_S}^2 \ {\rm sgn}(k).
\eeq
In other words, the anomalous radial energy flux is the product of this
angular momentum flux and the Doppler-shifted wave pattern speed
$\omega/m - \Omega$.  It therefore is identifiable as a true wave
energy flux.  (Turbulent energy and angular momentum fluxes, by way of
contrast, are related by a factor $\Omega$.) Its significance is that
in a $Q\sim1$ disk it will contribute to the total energy flux at a
level comparable to the stress tensor $W_{R\phi}$ if  $|\omega/m -
\Omega|/\Omega $ is of order unity.  The effect is to  prevent
self-gravitating disks from behaving like $\alpha$ disks; only if the
anomalous flux vanishes can a self-gravitating disk behave like a local
$\alpha$ disk.  The ``forbidden zone''of wave propagation near the
corotation point $\omega = m\Omega$ will display properties similar to
an $\alpha$ disk.  However, when a disk location undergoes forcing due
to a potential from a wave pattern rotating with a frequency
very different from the local rotation frequency, it will not behave
like an $\alpha$ disk.  Such a situation may occur, for example, when
an exterior disk is forced by the potential caused by a developing
central bar instability.

Energy can be exchanged between fluctuations and the differential
rotation of the disk; unlike angular momentum, it need not be conserved
in the noncircular motions.  In contrast to a turbulent $\alpha$ disk,
a self-gravitating disk can evolve by extracting energy from the
background shear and allocating it to the flucutations (wave energy)
without the need for mechanical energy dissipation.  This allows for
angular momentum transport with no associated {\em local\/} energy
losses.  Significant angular momentum transport of this type can occur
if a global nonaxisymmetric mode develops in an initially
gravitationally unstable disk.  Such a construction need not be merely
a transient initial condition.  Such features are semi-permanent,
slowly evolving as the disk background parameters change (Papaloizou \&
Savonije 1991, Papaloizou 1996, Laughlin \& Royczyska 1996). More
recently, careful analyses of self-gravitating disk simulations carried
out by Laughlin, Korchagin \& Adams (1997, 1998) clearly show angular
momentum transport produced by the onset of global nonaxisymmetric
instability and subsequent generation of a global wave pattern which
has extracted energy from the background shear.  Transport of this
type, which is seen to persist even after initial saturation, certainly
does not have the character of that exhibited by an $\alpha$ disk
(Laughlin \& Royczyska 1996).  These simulations, however, involve massive
disks (comparable stellar and disk masses); similar studies of 
lower mass Keplerian disks have yet to be done.

\subsection{Conditions Under Which Self-Gravity Leads to an $\bb{\alpha}$ Disk}

There is a local limit in which the nonlocal energy flux terms vanish
and eq.~(\ref{abc}) is recovered.  It occurs when the shearing box
limit is used to study self-gravitating disks.

The shearing box approximation is a standard approach to the dynamics
of thin disks, both self-gravitating (Goldreich \& Lynden-Bell 1965,
Julian \& Toomre 1966, Toomre 1981) and non self-gravitating
(Goldreich, Goodman, \& Narayan 1986; Hawley, Gammie, \& Balbus 1995).
In this scheme, the disk is divided into local Cartesian patches with
periodic boundary conditions being applied on their boundaries. Thus
one considers a small box in the disk, and sets up local corotating
coordinates corotating with the patch center.  Strictly periodic
boundary conditions are applied in the azimuthal direction, so that
$\phi$-averaging amounts to averaging over one azimuthal width of the
box.  However, in the radial direction, because of the presence of
large scale shear, periodicity is applied at boundary points which are
azimuthally separating from one another.  Related periodic points must
shear apart with time.  Thus, strict periodicity would hold only in
comoving, {\em shearing\/} coordinates.  Note that there is no
preferred location in this description so the box may be centered
anywhere (except, of course, the origin).

The significance of these boundary conditions is that they force the 
divergence of $\bb{F^*_E}$ to be zero when averaged over the box.  If 
$\Delta R$ is the radial extent of the box, and $R\pm\Delta R/2$
represent the outer ($+$) and inner ($-$) boundaries, then the integrated
box average leads to a term of the form
\beq
\left[ \left\langle  {\dd \Phi_S\over \dd R}{D\Phi_S\over D t}
\right\rangle_\phi \right]^{R+\Delta R/2}_{R-\Delta R/2},
\eeq
which must vanish.  The square bracket notation denotes a difference to
be taken between the upper and lower indicated locations.  The boundary
conditions force every fluid element on the inner edge to have a
corresponding partner on the outer edge, and the appearance of $D/Dt$,
rather than $\dd/\dd t$, ensures cancellation.  Were the partial time
derivative used, we would not be forced to this conclusion, because the
disk passes by ``faster'' at one of the boundaries compared with the
other.  If $\bb{F^*_E}$ vanishes (in this averaged sense),
we are lead to a standard $\alpha$ disk model.
Clearly, however, this conclusion is entirely driven by the choice of boundary
conditions.  Energy loss or gain from an evolving wave-like flux would be quite
incompatible with this type of periodicity.

There is no physical reason for the above boundary conditions to be
satisfied in the neighborhood of an arbitrarily chosen disk location.
Nevertheless, it is possible that there are circumstances under which
$\bb{F^*_E}$ may in effect vanish.  Disks evolving under the influence
of their own self-gravity tend to hover near the critical $Q=1$ level
(Laughlin \& Bodenheimer 1994).  WKB waves with radial wavenumber
$k_{crit} = \pi G \Sigma/c_S^2\sim 1/(QH)$ are neutrally stable
($\omega - m \Omega =0$); all other wavennumbers propagate.  While
ostensibly comparable in magnitude to $\bb{ {\cal F}}$, $\bb{F^*_E}$
may be smaller in a $Q=1$ disk.  The most responsive (dominant?) local
modes have $\omega -m\Omega\ll\Omega$, and this may be enough to
suppress $\bb{F^*_E}$.  The effectivenss of this process depends both
upon the disk's alacrity in maintaining $Q$ near unity, and upon the
shape of the wave power spectra.  Clearly, numerical simulations are
needed to resolve the question of whether $Q=1$ disks can be treated
within the $\alpha$ formalism.

Values of $Q$ near unity are favored, of course, because dropping below
this critical level results in vigorous dissipative shock heating,
raising the temperature and stabilizing.   Rising above $Q=1$ allows
the disk to cool and become destabilized (e.g., Sellwood \& Carlberg
1984).  The critical criterion also has some observational support
through the work of Kennicutt (1989), who has found that the gaseous
$Q$ value of active star-forming regions of disk galaxies is near
critical.  More generally, it is yet well understood under what
conditions heating and cooling will be able to regulate $Q$ efficiently
in disks, thereby allowing the use of a simplifying $\alpha$
formalism.

\section {Summary}

The dynamical foundations of viscous $\alpha$ disk models are
rooted in the correlated fluctuations which create the underlying
turbulent stresses.  In this paper, we have shown that the mean flow
dynamics of MHD turbulence follows the $\alpha$ prescription, and in
particular that the disk energy dissipation rate is always give by
eq.~(\ref{abc}), even if the disk is evolving.  The local character of
MHD disturbances is itself rooted in the flux freezing equation, which
forces local dissipation of the magnetic field in turbulent flow,
analogous to vorticity dynamics in an unmagnetized shear layer.

The mean flow dynamics of a self-gravitating disk in general cannot be
described so simply.  Classical viscous disk theory requires a simple
restrictive form for the mean momentum and energy fluxes
(eqs.~[\ref{momflux}] and [\ref{eflux}]); neither can depend upon
transport properties other than $\langle u_R \rangle_\rho$ and
$W_{R\phi}$.  The energy flux of self-gravitating disks is not
reducible to a superposition of these quantities.  Instead, what we
refer to as anomalous flux terms are present.  These terms allow
self-gravitating disturbances (not necessarily of WKB form) to
propagate nonlocally in the disk via the perturbed gravitational
potential; a viscous disk cannot communicate with itself in a similar
fashion.  The angular momentum flux (strictly conserved) in a
self-gravitating disk has the same canonical form it has in a non
self-gravitating disk, proportional to $W_{R\phi}$; the energy flux is
fundamentally different.

In a non self-gravitating thin disk, wave energy transport depends upon
terms in the flux which, while formally present, are small by order
$H/R$.  In a self-gravitating disk, the additional (non-pressure) terms
that are present in the energy flux couple directly to the differential
rotation of the disk, as does $W_{R\phi}$.  This additional coupling
means in effect
that transport becomes global on rotational time scales.  Over
similar times without self-gravity,
the domain of wave influence is
restricted to the vertical scale-height $H$.

Shearing box simulations of self-gravitating disks employ boundary
conditions which force local behavior, and inevitably must give rise to
an $\alpha$ disk.  Because self-gravity is intrinsically nonlocal in
its manifestations, analyzing transport phenomena in self-gravitating
disks within the shearing box formalism may be misleading.   On the
other hand, it is possible that critical $Q\simeq 1$ disks will be
dominated by wavenumbers for which $|\omega/m -  \Omega|/\Omega$ is
small, in which case $\alpha$ modeling might be a fair phenomenological
description.  To date however, global numerical simulations of massive
self-gravitating disks, do not seem to lend themselves readily to an
$\alpha$ formalism.  Whether the same is true for self-gravitating
disks much less massive then their central stars is not yet known.

\bigskip

We thank J.~Hawley for useful discussions.  S.A.B. acknowledges support from
NASA grants NAG5-3058, and NAG5-7500, and NSF grant AST-9423187.
J.C.B.P. acknowledges support from PPARC Grant GR/H/ 09454.

\clearpage
\begin{center}
{\bf References}
\end{center}
\def\refindent{\par\penalty-100\noindent\parskip=4pt plus1pt
	       \hangindent=3pc\hangafter=1\null}

\refindent Balbus, S.~A., \& Hawley, J.~F. 1991, ApJ, 376, 214
\refindent Balbus, S.~A., \& Hawley, J.~F. 1998, Rev Mod Phys, 70, 1
\refindent Binney, J., \& Tremaine, S. 1987, Galactic Dynamics
(Princeton: Princeton University Press)
\refindent Goldreich, P., Goodman, J., \& Narayan, R. 1986, MNRAS, 221, 339
\refindent Goldreich, P., \& Lynden-Bell, D. 1965, MNRAS, 130, 125
\refindent Goldreich, P., \& Tremaine, S.~D. 1979, ApJ, 233, 857 (GT)
\refindent Hawley, J.~F., Gammie, C.~F., \& Balbus, S.~A. 1995, ApJ, 440, 742
\refindent Julian, W.~H., \& Toomre, A. 1966, ApJ, 146, 810
\refindent Kennicutt, R.~C. 1989, ApJ, 344, 685
\refindent Krause, F., \& R\"adler, K.-H. 1980, Mean-Field Magnetohydrodynamics and 
Dynamo Theory, (Oxford: Pergammon)
\refindent Landau, L.~D., \& Lifschitz, E.~M. 1959, Fluid Mecahnics, (Oxford: Pergammon)
\refindent Laughlin, G., \&  Bodenheimer, P. 1994, ApJ, 436, 335
\refindent Laughlin, G., Korchagin, V., \& Adams, F.~C. 1997, ApJ, 477, 410
\refindent Laughlin, G., Korchagin, V., \& Adams, F.~C. 1998, ApJ, 504, 945
\refindent Laughlin, G., \& Royczyska, M. 1996, 456, 279
\refindent Lighthill, J. 1978, Waves in Fluids, (Cambridge: Cambridge University
Press)
\refindent Lin, C.~C., \& Shu, F.~H. 1966, ApJ, 55, 229
\refindent Lynden-Bell, D., \& Kalnajs, A.~J. 1972, MNRAS, 157, 1
\refindent Lynden-Bell, D., \& Pringle, J.~E. 1974, MNRAS, 168, 603 
\refindent Papaloizou, J.~C. 1996, 
in Gravitational Dynamics, ed. O. Lahav, E. Terlevich, \& R. Terlevich 
(Cambridge: Cambridge Univ. Press), p. 119
\refindent Papaloizou, J.~C., \& Savonije, G.~J. 1991, MNRAS, 248, 353
\refindent Pringle, J.~E. 1981, ARAA, 19, 137   
\refindent Sellwood, J.~A., \& Carlberg, R.~G. 1984, ApJ, 282, 61
\refindent Shakura, N.~I., \& Sunyaev, R.~A. 1973, AA, 24, 337
\refindent Toomre, A. 1964, ApJ, 139, 1217
\refindent Toomre, A. 1981, in The Structure and Evolution of Normal Galaxies, eds.
S.~M. Fall and D.~Lynden-Bell (Cambridge: Cambridge University press), p. 111
\refindent Welsh, W.~F., Wood, J.~H., \& Horne, K. 1996, in Cataclysmic
Variables and Related Objects; Proceedings of the 158th Colloquium of the
International Astronomical Union, eds. A. Evans and J.~H. Wood
(Dordrecht: Kluwer), p. 29

\end {document}